\documentclass[aps,prl,twocolumn,superscriptaddress,nofootinbib,longbibliography]{revtex4-2}

\usepackage{amsmath,amssymb,amsfonts}
\usepackage{graphicx}
\usepackage{hyperref}
\usepackage{xcolor}
\usepackage{bm}
\usepackage{subcaption}

\begin{document}

\title{Phantom crossing from the Standard Model and General Relativity}

\author{Jinglong Liu}
\email{jinglong\_liu@sjtu.edu.cn}
\affiliation{Tsung-Dao Lee Institute, 1 Lisuo Road, Shanghai, 201210, China}
\affiliation{School of Physics and Astronomy, Shanghai Jiao Tong University, 800 Dongchuan Road, Shanghai 200240, China}

\author{Tucker Manton}
\email{tucker\_manton@ucas.ac.cn}
\affiliation{School of Fundamental Physics and Mathematical Sciences, Hangzhou Institute for Advanced Study,\\
\phantom{}\hspace{0.5cm} University of Chinese Academy of Sciences (HIAS-UCAS), Hangzhou, 310024, China}

\author{Yifu Cai}
\email{yifucai@ustc.edu.cn}
\affiliation{Department of Astronomy, School of Physical Sciences, University of Science and Technology of China, Hefei 230026, China}
\affiliation{CAS Key Laboratory for Research in Galaxies and Cosmology, School of Astronomy and Space Science, 
University of Science and Technology of China, Hefei 230026, China}

\author{Antonino Marcian\`o}
\email{marciano@fudan.edu.cn}
\affiliation{Department of Physics \& Center for Astronomy and Astrophysics, Fudan University, 200433 Shanghai, China}
\affiliation{Laboratori Nazionali di Frascati INFN, Frascati (Rome), Italy, EU}

\date{\today}

\begin{abstract}
\noindent 
%Recent analyses combining DESI baryon acoustic oscillation measurements with cosmic microwave background and Type-Ia supernova observations have renewed interest in evolving dark energy and possible crossings of the phantom divide. We propose a late-time cosmological scenario in which dark energy originates from a curvature-triggered phase transition of a fermion condensate. The condensate develops dynamically once the Ricci scalar falls below a critical threshold, generating a vacuum component whose equation of state exhibits an intrinsic phantom singularity. We show that nonlinear structure formation simultaneously induces a geometric contribution described by Buchert averaging. Although this backreaction component is insufficient to explain cosmic acceleration by itself, it regularizes the condensate singularity and produces a smooth phantom-crossing evolution. Matching the observed dark-energy abundance predicts a transition redshift $z_c\simeq2.45$, providing a falsifiable target for future surveys.
%
Suggestions of a late-time phantom crossing from DESI baryon acoustic oscillation measurements, combined with cosmic microwave background and Type-Ia supernova observations, have renewed interest in non-standard dark energy models. In this Letter, we propose a new realization of a low redshift phantom crossing using only well understood ingredients from fermion condensation and general relativity. Our construction relies on the interplay between effective quintom-like dark energy and  backreaction from non-linear structure formation. The full cosmological evolution is consistent with $\Lambda$CDM until low redshift, where the backreaction starts to become prominent and a phase transition occurs. At this point, the total dark energy equation of state first crosses $w_{\text{DE}}=-1$ from above. For benchmark values of the backreaction energy density, a second crossing from below occurs at a lower redshift, as suggested by recent observation.  Fitting our theoretical model with CPL parametrization, we find the result is consistent with the combined DESI+CMB+SNIa data analysis with DESY5, PantheonPlus, and Union3 datasets. Specifically, for backreaction density parameter $\Omega_{\rm BR}(z=0) = 0.0572$, the CPL parametrization gives the phantom crossing redshift $z_* \simeq 0.35$, $w_0 \simeq -0.76 $ and $w_a \simeq -0.93$.
\end{abstract}

\maketitle

\noindent 
{\it Introduction. ---} The physical origin of dark energy remains one of the most profound open questions in modern cosmology. Since the discovery of cosmic acceleration through Type-Ia supernova observations \cite{Riess:1998cb,Perlmutter:1998np}, a broad range of measurements, including the cosmic microwave background (CMB), baryon acoustic oscillations (BAO), weak lensing, and large-scale structure surveys, have converged toward a cosmological model in which approximately $70\%$ of the present energy density is stored in a dark component with negative pressure \cite{Planck:2018vyg}. The simplest realization of this phenomenon is provided by a positive cosmological constant $\Lambda$, corresponding to an equation-of-state parameter $w=-1$. Despite its remarkable observational success, the cosmological constant suffers from severe theoretical challenges, including the vacuum-energy hierarchy problem and the coincidence problem, which have motivated extensive investigations of dynamical dark-energy scenarios \cite{Copeland:2006wr,Tsujikawa:2013fta}.

A particularly intriguing possibility is that dark energy evolves with cosmic time. Recent analyses combining DESI BAO measurements with CMB and Type-Ia supernova datasets have reported a mild but persistent preference for departures from the standard $\Lambda$CDM scenario \cite{DESI:2024mwx}. Similar trends have previously emerged in eBOSS analyses \cite{eBOSS:2020yzd} and have recently been reinforced through combinations involving DESY5 supernovae \cite{DES:2026jmi}, PantheonPlus \cite{Scolnic:2021amr}, and Union3 compilations. Although none of these results independently constitute definitive evidence against $\Lambda$CDM, they collectively suggest that an evolving dark-energy sector deserves serious theoretical consideration.

The observational preference is often expressed through the Chevallier--Polarski--Linder (CPL) parameterization \cite{Chevallier:2000qy,Linder:2002et}, $w(a)=w_0+w_a(1-a)$, 
for which the cosmological constant corresponds to $(w_0,w_a)=(-1,0)$. Interestingly, current best-fit regions frequently extend into the phantom domain $w<-1$. Specifically, using the joint analysis of DESI BAO+CMB+DES Y5 data, the DESI collaboration has reported  $w_0=-0.727\pm 0.067$ and $w_a=-1.05^{+0.31}_{-0.27}$ at $3.9\sigma$ significance \cite{DESI:2024mwx}. By taking the center values of the DESI result, we deduce that there was a possible phantom crossing around redshift $z\sim 0.35.$

 Such a crossing is notoriously difficult to realize in a theoretically consistent framework. General no-go theorems show that canonical single-field models cannot smoothly cross the phantom divide without encountering singular behavior or gradient instabilities \cite{Vikman:2004dc}. Conventional quintom constructions evade this difficulty by introducing additional degrees of freedom, typically involving phantom fields with negative kinetic energy \cite{Feng:2004ad,Cai:2009zp}. While phenomenologically successful, these constructions often suffer from vacuum instabilities and raise concerns regarding the fundamental consistency of the underlying theory.

In parallel, a different line of investigation has explored the possibility that dark energy may emerge from condensate phenomena rather than from elementary scalar fields. Fermion-condensate models, inspired by Nambu--Jona-Lasinio dynamics and gravitationally-induced four-fermion interactions, provide a particularly attractive realization of this idea \cite{Addazi:2017qus,Alexander:2025whu,Alexander:2026jlo}. In these scenarios, a composite order parameter dynamically develops a vacuum expectation value and acts as an effective dark-energy degree of freedom. Such constructions naturally avoid the introduction of ad hoc fundamental scalars, use only Standard Model degrees of freedom, and establish a direct connection between cosmology and microscopic fermionic dynamics.

In this Letter, we combine these ideas with the physics of nonlinear structure formation. A curvature-induced cosmological phase transition occurs at late times; the Ricci scalar falls below a critical threshold, triggering a nonzero solution for the fermion condensate order parameter $A$, at which moment the condensate's vacuum energy becomes cosmologically relevant. Remarkably, quantitative analysis indicates that the phase transition occurs in the late-time universe, coinciding with the onset of non-linear structure formation. In this regime, an effective dark-energy component emerges from the averaged local expansion rate, identified with Buchert backreaction \cite{Buchert:1999er,Buchert:1999mc,Buchert:2006ya,Galoppo:2026rin}. This backreaction dynamically elevates the dark-energy EoS, driving the system toward and across the phantom divide during the transition. After the transition, the EoS is below -1 and asymptotically approaches $w=-1$ in the far future, while allowing for a second phantom crossing near the present epoch.

These features lead to a distinctive late-time cosmological history that can be directly confronted with current DESI+CMB+SN observations and tested by future surveys such as Euclid \cite{Euclid:2019clj}, Roman \cite{Eifler:2020vvg}, and the LSST at the Rubin Observatory \cite{LSST:2008ijt} .\\

\noindent 
{\it Curvature-Triggered Condensation. ---} The effective low-energy dynamics of the condensate order parameter $A$ is described by \cite{Alexander:2025whu,Alexander:2026jlo}
\begin{equation}
V_{\rm eff}(A)
=
V_0
-\frac12 \delta m^2 A^2
+\frac{RL}{96\pi^2}A^2
+\frac{L}{4\pi^2}A^4 ,
\label{eq:Veff}
\end{equation}
where $R$ denotes the Ricci scalar and $L\equiv
\ln\!\left({\Lambda^2}/{A^2}\right)$. The order parameter $A$ is explicitly related to the fermion bilinear as $m^2A=\langle\bar{\psi}\psi\rangle+\langle\bar{\psi}\gamma^5\psi\rangle.$ The effective mass is
\begin{equation}
\mu_{\rm eff}^2
=
-\delta m^2
+\frac{RL}{48\pi^2}.
\end{equation}
At early times the large positive curvature maintains the symmetric phase $A=0$. Symmetry breaking occurs once
\begin{equation}
R<R_c,
\qquad
R_c
=
\frac{48\pi^2}{L}\delta m^2 .
\label{eq:Rc}
\end{equation}
The condensate expectation value then becomes
\begin{equation}
A^2(R)
=
\frac{\pi^2}{L}\delta m^2
\left(
1-\frac{R}{R_c}
\right).
\end{equation}

Assuming matter domination around the transition epoch, $R(a)=R_0a^{-3}$, with $R_0=3\Omega_{m0}H_0^2$, the critical scale factor is  
\begin{equation}
a_c
=
\left(
\frac{R_0}{R_c}
\right)^{1/3}
=
\left(
\frac{L\Omega_{m0}H_0^2}{16\pi^2\delta m^2}
\right)^{1/3},
\qquad
1+z_c=a_c^{-1}.
\label{eq:ac}
\end{equation}

For $z<z_c$, the condensate energy density is
\begin{equation}
\rho_A(z)
=
\rho_{\rm vac}
\left[
1-
\left(
\frac{1+z}{1+z_c}
\right)^3
\right]^2 \,.
\label{eq:rhoA}
\end{equation}
Using energy conservation $\dot\rho_A+3H(\rho_A+p_A)=0$, one obtains
%
%\textcolor{red}{[AM: There should not be a minus sign after the equality below?]}
%
\begin{equation}
w_A(z)
= -
\frac{
(1+z_c)^3+(1+z)^3
}{
(1+z_c)^3-(1+z)^3
}.
\label{eq:wA}
\end{equation}
Equation~(\ref{eq:wA}) predicts a future de-Sitter attractor but also reveals an intrinsic phantom singularity at $z=z_c$. However, the singularity can be removed when constant $V_0$ is recovered.\\

\noindent 
Evading the cosmological constant problem. --- The phase transition scale is fixed by matching the condensate abundance to the observed dark-energy density. The bare asymptotic depth of the condensate is $\rho_{\rm vac}^{\rm bare}=\frac{\pi^2}{4L}\delta m^4$. Using Eq.~(\ref{eq:ac}) to eliminate $\delta m$ provides us with
\begin{equation}\label{delta}
\delta m^2
=
\frac{L\Omega_{m0}H_0^2}{16\pi^2}
(1+z_c)^3 .
\end{equation}
The ultraviolet scale that suppresses the microscopic four-fermion interaction (responsible for the formation of the cosmological scalar condensate) shall be distinguished from the infrared scale associated with the condensate gap. Collider searches constrain local contact interactions expressed by $\Lambda_{\rm contact}^{-2}
(\bar\psi\Gamma\psi)
(\bar\psi\Gamma\psi)$, requiring $\Lambda_{\rm contact}$ to lie above the multi-TeV range. These limits apply to the ultraviolet completion of the theory and do not directly constrain the infrared quantity $\delta m^2=\Lambda^2/\pi^2-m^2$, which measures the distance from the Nambu--Jona-Lasinio critical surface. Differently the mass scale $\delta m\sim H_0$, which has cosmological relevance, should not be interpreted as the microscopic interaction scale, but rather as the renormalized infrared gap generated near criticality.

For $L\simeq130$ and $\Omega_{m0}\simeq0.315$, %this 
Eq.~\eqref{delta} 
implies
\begin{equation}
\rho_{\rm vac}^{\rm bare}
\simeq
0.001283\,
(1+z_c)^6
H_0^4 .
\end{equation}

Since the condensate is triggered by the global curvature scale, we assume a global holographic enhancement by the number of horizon degrees of freedom, i.e. ${\cal N} \simeq {M_{\rm Pl}^2}/{H_0^2}$ such that $\rho_{\rm vac}^{\rm eff}={\cal N}\rho_{\rm vac}^{\rm bare}$. This enhancement can be naturally interpreted as the macroscopic imprint of the cosmological horizon on the condensate. The infrared spectrum of the fermion condensate is therefore determined by the finite size of the observable Universe, the largest physical wavelength of which is order $H_0^{-1}$. Thus $\delta m$ represents an emergent infrared mass rather than a microscopic coupling. The phase transition is driven by the evolution of the cosmological infrared cutoff. This naturally explains why the condensate becomes dynamically relevant only at late times. The effective macroscopic vacuum depth hence turns out to be
\begin{equation}
\rho_{\rm vac}^{\rm eff}
\simeq
0.001283\,
(1+z_c)^6
M_{\rm Pl}^2H_0^2 .
\label{eq:rhoeff}
\end{equation}
Evaluating Eq.~(\ref{eq:rhoA}) today entails
\begin{equation}
\rho_A(0)
=
\rho_{\rm vac}^{\rm eff}
\left[
1-(1+z_c)^{-3}
\right]^2 .
\end{equation}
Thus, one can find
\begin{equation}
\Omega_{A0}
=
\frac{\rho_A(0)}
{3M_{\rm Pl}^2H_0^2}
=
\frac{0.001283}{3}
\left[
(1+z_c)^3-1
\right]^2 .
\label{eq:omegazc}
\end{equation}

\noindent 
{\it Geometric Backreaction. ---} The transition occurs during the epoch of nonlinear structure formation. This regime allows us to employ Buchert averaging \cite{Buchert:1999er,Buchert:1999mc,Buchert:2006ya}, including the kinematical backreaction $\mathcal Q_{\mathcal D}
=
\frac23
\left(
\langle\theta^2\rangle_{\mathcal D}
-
\langle\theta\rangle_{\mathcal D}^{\,2}
\right)
-
2\langle\sigma^2\rangle_{\mathcal D}$ and a curvature average $\langle R\rangle_\mathcal{D}$. In $\mathcal{Q}_{\mathcal D}$, the first term is the variance of expansion rate, while the second term is the average shear. From the averaged Raychaudhuri equation, the corresponding effective fluid is characterized by
\begin{align}
\rho_{\rm BR}
&=
-\frac{1}{16\pi G}
\left(
\mathcal Q_{\mathcal D}
+
\langle\mathcal R\rangle_{\mathcal D}
\right),
\\
p_{\rm BR}
&=
-\frac{1}{16\pi G}
\left(
\mathcal Q_{\mathcal D}
-\frac13
\langle\mathcal R\rangle_{\mathcal D}
\right).
\end{align}
Relativistic simulations \cite{Buchert:2006ya,Bentivegna:2015flc} and recent data analysis from Cosmicflows-4++ \cite{Galoppo:2026rin} typically find $\mathcal Q_{\mathcal D} \ll |\langle\mathcal R\rangle_{\mathcal D}|$, so that $w_{\rm BR} \simeq -\frac13$. This component cannot drive cosmic acceleration by itself. However, its presence regularizes the condensate transition. We define $\rho_{\rm DE}=V_0 + \rho_A+\rho_{\rm BR}$ and $p_{\rm DE}=-V_0+p_A+p_{\rm BR}$, 
and hence 
\begin{equation} \label{eq:weff}
w_{\rm eff} =\frac{p_{\rm DE}}{\rho_{\rm DE}} \,.
\end{equation}
The geometric contribution stabilizes the denominator of Eq.~(\ref{eq:weff}), eliminating the condensate divergence and yielding a finite phantom-crossing evolution. \\

The parameters $V_0, \Omega_{\rm BR0}$ and $z_c$ satisfy two constraint equations from observations:
the total dark energy density and the  acceleration $q = 0$ at $z_{acc} \simeq 0.65$ \cite{Farooq:2016zwm}, characterizing the transition from matter domination to  accelerate expanding phase. The two equations are:
\begin{equation}
    V_0 + \rho_{\rm vac} \left[ 1 - \frac{1}{(1+z_c)^3} \right]^2 + \rho_{\rm BR0} = \rho_{\Lambda0}^{\rm obs} \simeq 0.685\rho_{\rm crit},
    \label{eq:constraint_budget}
\end{equation}
and
\begin{equation}
    \rho_{m0}(1+z_{\rm acc})^3 - 2V_0 + \rho_A(z_{\rm acc}, z_c) + 3p_A(z_{\rm acc}, z_c) = 0,
    \label{eq:constraint_acc}
\end{equation}
where $\rho_{\rm BR0} = -6\Omega_{\rm BR0} \mathcal{H}_\mathcal{D}^2 \approx \langle R\rangle_{\mathcal{D}}$ at present \cite{Galoppo:2026rin}. Solving eqs.\eqref{eq:constraint_budget} and \eqref{eq:constraint_acc} with specific $\Omega_{\rm BR0}$ determines $z_c$ and $V_0$.\\

\noindent
{\it Phantom Crossing and Data Comparison. ---} The phenomenology of the model is governed by the interplay between two physically distinct contributions to the dark energy sector. On the one hand, the fermion condensate generates a vacuum component whose energy density grows after the curvature-induced phase transition. On the other hand, nonlinear structure formation induces an effective geometric backreaction described by the Buchert formalism. The backreaction energy density parameter is chosen below $\Omega_{\rm BR0} \lesssim 0.065$, in alignment with the value given by Ref.~\cite{Galoppo:2026rin}. Since the backreaction cannot be traced back to the early universe, we assume an evolution function \begin{equation}
    S_{\rm NL}(z) = \frac{1}{2} \left[ 1 - \tanh\left(\frac{z - z_{\rm nl}}{\delta_{\rm nl}}\right) \right],
\end{equation}
where $z_{\rm nl} \sim 1.5$ denotes the characteristic onset of non-linear structure formation and $\delta_{\rm nl}$ represents its duration, which we take to be $\delta_{\rm nl}\sim 0.5$. The regularized macroscopic geometric fluid is thus parameterized as:
\begin{equation}
    \Omega_{\rm BR}(z) = \Omega_{\rm BR0} (1+z)^2 \frac{S_{\rm NL}(z)}{S_{\rm NL}(0)}.
\end{equation}

\begin{figure}[t]
\centering
\includegraphics[width=\columnwidth]{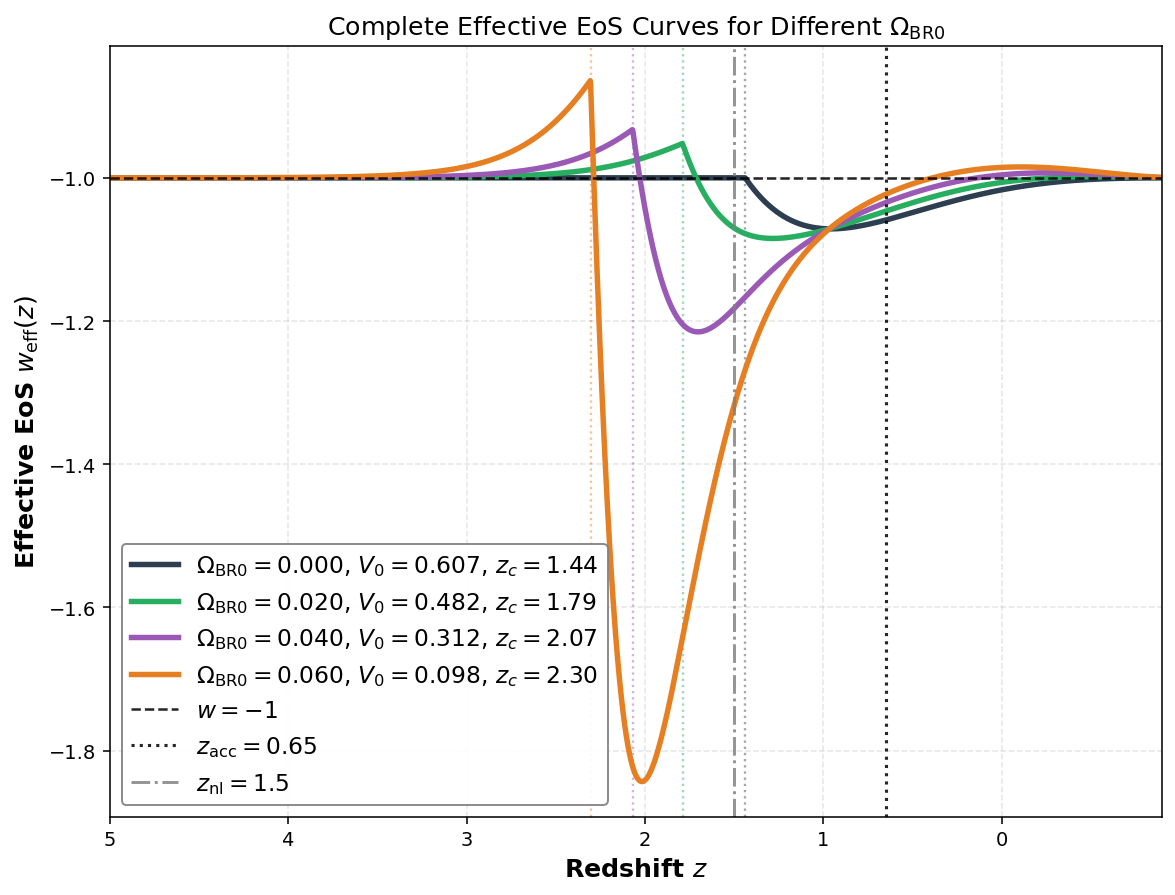}
\caption{
Evolution of the effective dark-energy equation of state for several transition redshifts. The phase transition generates a transient phantom excursion followed by relaxation toward the cosmological-constant limit.
}
\label{fig:eos}
\end{figure}

The resulting evolution of the effective equation-of-state parameter is shown in Fig.~\ref{fig:eos}. The backreaction lifts the EoS to $w>-1$. Then, after the phase transition occurring at a redshift $z_c$, the condensate rapidly develops a negative vacuum pressure and triggers a phantom crossing, while its energy density is still building up toward the asymptotic vacuum value. For larger $\Omega_{\rm BR0}$, a second phantom crossing can occur, as illustrated by the orange curve in Fig. \ref{fig:eos}. \\

The phantom excursion should not be interpreted as the manifestation of a fundamental ghost degree of freedom. Rather, it emerges dynamically from the non-equilibrium relaxation of the condensate order parameter following the phase transition \cite{Lulli:2021ricciflow}. As the condensate approaches its asymptotic vacuum configuration, the effective pressure gradually stabilizes and the equation of state evolves back toward the cosmological constant attractor.

%A particularly noteworthy feature of the evolution is the appearance of a pronounced ``phantom valley'' immediately after the transition. In the absence of geometric backreaction, the condensate contribution alone would develop a singular behavior, corresponding to the divergence discussed surrounding Eq.~(\ref{eq:wA}). The geometric sector regularizes this divergence by providing a finite positive contribution to the total energy density. Consequently, the singularity is replaced by a smooth minimum whose depth and duration are controlled by the transition redshift $z_c$ and by the relative importance of the backreaction component.

%Fig.~\ref{fig:eos} displays the evolution for several representative values of $z_c$. Larger values of the transition redshift correspond to an earlier activation of the condensate and therefore allow a longer relaxation period before the present epoch. As a consequence, the present-day equation of state remains extremely close to $w=-1$. Conversely, earlier transitions generate a more pronounced phantom behavior that remains visible in the low-redshift observational window. The value preferred by the dark-energy abundance constraint, $z_c\simeq2.45$, lies precisely in the regime where the phantom excursion is substantial enough to leave an observable imprint while still converging to a nearly cosmological-constant behavior today.

To compare the model with observations, we reconstruct its low-redshift evolution using the CPL parameterization $w(a)= w_0+w_a(1-a)$, which provides a convenient effective description in the interval $0\le z \lesssim 2$. Although the underlying evolution predicted by the condensate-backreaction system is intrinsically non-CPL, the parameterization offers a useful bridge to current observational analyses and allows a direct comparison with the dark-energy reconstructions reported by DESI and related surveys \cite{DESI:2025fxa,DESI:2025zgx}.

\begin{figure}[t]
\centering
\includegraphics[width=\columnwidth]{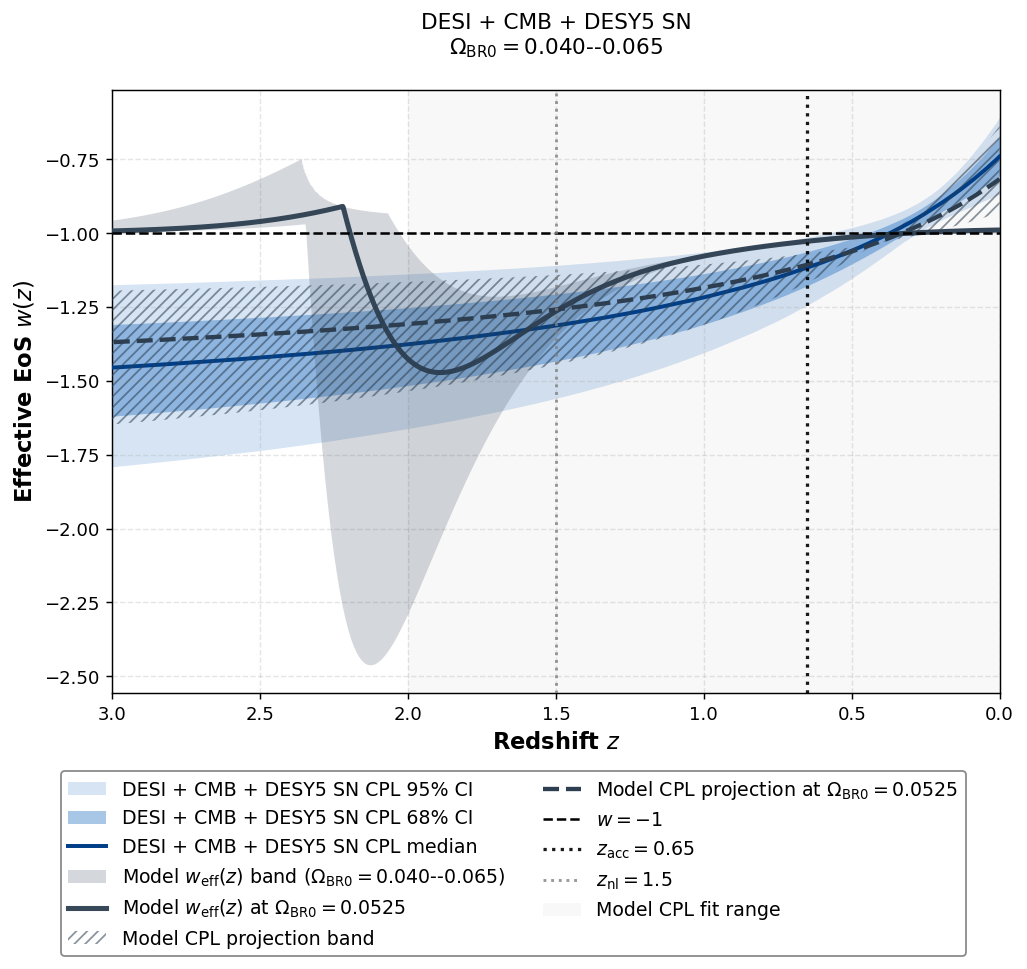}
\caption{
Comparison between the effective dark-energy evolution and DESI+CMB+DESY5 constraints.
}
\label{fig:desy5}
\end{figure}

Fig.~\ref{fig:desy5} compares the reconstructed evolution of the model with the joint DESI+CMB+DESY5 analysis. The theoretical trajectory lies within the region preferred by the observational contours and reproduces the tendency toward an evolving equation of state. In particular, the model naturally predicts a phase in which the effective equation of state falls below the cosmological-constant boundary before evolving back toward $w=-1$. This behavior closely mirrors the qualitative trend emerging from recent DESI analyses.

\begin{figure}[t]
    \centering

    \begin{subfigure}{0.9\columnwidth}
        \centering
        \includegraphics[width=\linewidth]{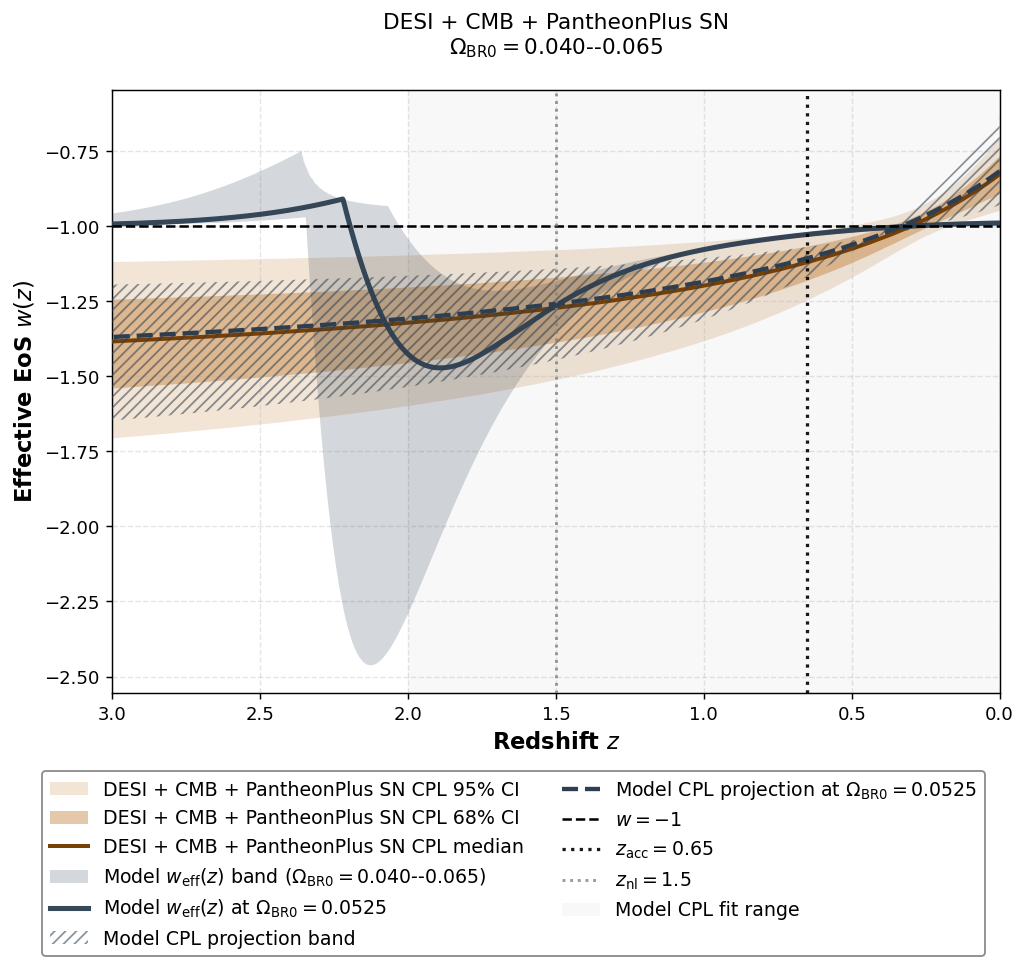}
    \end{subfigure}

    \vspace{0.5em}

    \begin{subfigure}{0.9\columnwidth}
        \centering
        \includegraphics[width=\linewidth]{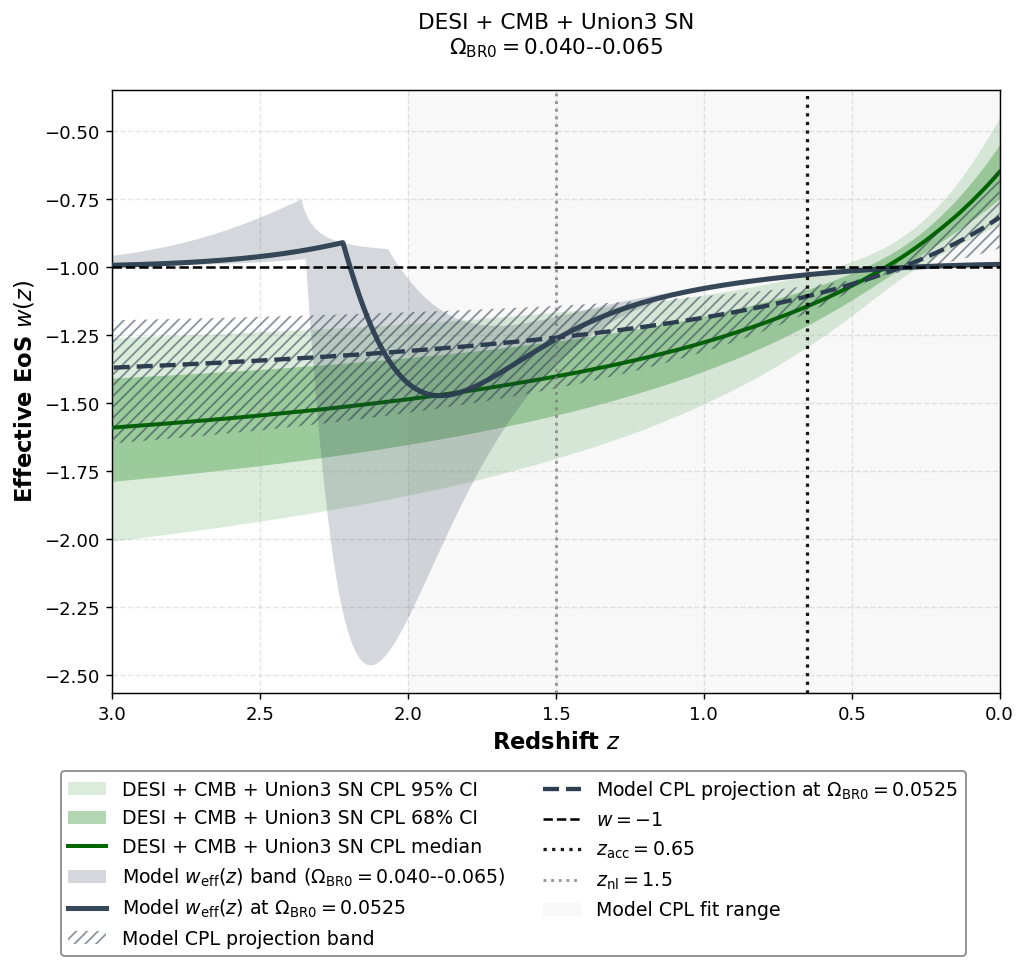}
    \end{subfigure}

    \caption{Comparison between the theoretical model, its CPL parametrization, and observational constraints from DESI+CMB+PantheonPlus and DESI+CMB+Union3.}
    \label{fig:sn}
\end{figure}

To assess the robustness of this conclusion, we further compare the model with two independent supernova compilations. Fig.~\ref{fig:sn} shows the corresponding results obtained using the PantheonPlus and Union3 datasets in combination with DESI and CMB. Despite differences in the reconstruction methodology and data selection, both analyses display a qualitatively similar preference for an evolving dark-energy sector. The predicted trajectory remains compatible with the preferred regions of parameter space and consistently reproduces the characteristic tendency toward a transient phantom regime.

Taken together, Figs.~\ref{fig:eos}--\ref{fig:sn} suggest a coherent physical picture. The dark-energy sector undergoes a late-time phase transition, generating a transient phantom epoch that is subsequently regularized by geometric backreaction. The resulting evolution reproduces the main qualitative features currently favored by DESI+CMB+SN analyses while predicting a specific transition scale $z_c$ and a cosmological constant $V_0$ when $\Omega_{\rm BR0}$ is fixed. Since the transition occurs in the redshift interval $1.44\lesssim z\lesssim 2.39$, where forthcoming observations from Euclid, Roman, Rubin, and future DESI releases will substantially improve precision, the model offers a clear observational target capable of distinguishing it from both $\Lambda$CDM and conventional quintom scenarios. Finally, we emphasize, when $\Omega_{\rm BR0} \simeq 0.0572$, the phantom crossing in CPL parametrization happens at $z_* \simeq 0.35$, $w_0 = -0.758 \pm 0.013$ and $w_a = -0.931 \pm 0.027$., matching very well with the observational constraints.
\vspace{0.3cm}

\noindent 
{\it Conclusions. ---} We have proposed a late-time dark-energy scenario in which cosmic acceleration emerges from a curvature-triggered phase transition of a fermion condensate. In this framework, the condensate remains frozen throughout most of cosmic history and becomes dynamically active only when the background Ricci curvature falls below a critical threshold. 

A central result of this work is that the phase transition happens when the Universe is treated as a genuinely inhomogeneous system and the nonlinear structure formation generates a geometric backreaction.  Although the backreaction is subdominant in the overall energy budget, it provides a nontrivial contribution to the effective dark-energy sector. The interplay between the microscopic condensate dynamics and the macroscopic geometric response gives rise to a smooth phantom-crossing evolution without introducing fundamental phantom degrees of freedom, higher-derivative operators, or modifications to general relativity.

The resulting cosmological history exhibits several distinctive features. First, the model predicts a late-time phase transition at a redshift $z_c \in [1.44, 2.39]$, which is determined by the observed dark-energy abundance rather than imposed phenomenologically. Second, the effective equation of state undergoes a transient excursion into the phantom regime before relaxing toward a de-Sitter attractor with $w \rightarrow -1$, for $z\rightarrow -1$. Third, the predicted evolution qualitatively reproduces the behavior currently favored by several combinations of DESI, CMB, and Type-Ia supernova datasets.

More broadly, our results suggest that phantom-crossing behavior need not originate from fundamental ghost fields or exotic modifications of gravity. Instead, it may arise as an emergent phenomenon associated with the interplay between microscopic vacuum physics and the nonlinear evolution of cosmic geometry\footnote{Some  proposals have entertained the interesting possibility that evolving dark energy can be associated with the QCD vacuum dynamically responding to cosmological expansion \cite{VanWaerbeke:2025shm}. This idea was recently shown to be consistent with DESI in \cite{Lee:2026yzs}. }. In this sense, the phantom phase is not a fundamental property of the underlying theory, but rather a transient manifestation of a late-time non-equilibrium transition in the dark energy sector.

Future observations will provide a direct test of this picture. The predicted transition occurs within the redshift interval currently explored by DESI and will be probed with significantly greater precision by forthcoming surveys such as Euclid, Roman, and Rubin. Measurements of the expansion history, growth rate of structure, and possible deviations from the standard CPL reconstruction in the range $1\lesssim z\lesssim3$ will therefore offer a critical opportunity to distinguish this scenario from both $\Lambda$CDM and conventional quintom models.

In a companion paper \cite{Liu:2026}, we elaborate on many aspects of the late universe features of the model that we have omitted here in this Letter. This includes examining the deep theoretical connection between the emergence of the fermion vacuum dynamics, holographic regularization, and the cosmological constant problem. We further explore the consequences of our construction for early universe physics, where we show that a related out-of-equilibrium ghost mechanism induces a phantom crossing during inflation. The early and late universe behaviors of the model are theoretically and phenomenologically consistent, and provide a comprehensive cosmological history.

If confirmed, the framework presented here would point toward a novel connection between fermion-condensate vacuum physics, nonlinear gravitational dynamics, and the origin of cosmic acceleration, suggesting that the observed evolution of dark energy may ultimately be a consequence of a late-time phase transition in the vacuum structure of the Universe.

\vspace{0.2 cm}

\noindent\textit{Acknowledgments. ---} The authors thank Elisa Ferreira and Roman Pasechnik for useful discussions. The work of J.L. is supported in part by the National Key R\&D Program of China No. 2024YFC2207700.

\bibliographystyle{apsrev4-2}
\bibliography{References_PRL}

\end{document}